# Analyzing new planetary systems at school: Applications of Newton's Universal Gravitation and Kepler's Third Law

R. Montecinos[1,2,3], C. Hernández[1,2,3], I. Fuentes-Morales[1,2,3], F. Alarcón[2], I. Benito[2], L. Laroze[4] & S. Pérez[1,2,3]

(1) Universidad de Santiago de Chile, Santiago, Chile.
(2) Center for Interdisciplinary Research in Astrophysics and Space Sciences, CIRAS, Chile.
(3) Millennium Nucleus on Young Exoplanets and their Moons, YEMS, Chile.
(4) Universidad Técnica Federico Santa María, Santiago, Chile.

Email: ruben.montecinos@usach.cl

## Introduction

As scientific knowledge expands, high school science education may not always keep pace with the latest advancements in astrophysics. A solid scientific education is crucial for preparing students for 21st-century challenges[1]. However, science education often focuses narrowly on specific content, neglecting frontier scientific research. To address this, a teaching sequence was developed in Chile using real exoplanet data from the Open Exoplanet Catalog and NASA's Eye on Exoplanets webpage. This integrates cutting-edge astrophysical concepts into classroom discussions. Analyzing this data prompts students to discuss how Newton's law of universal gravitation and Kepler's third law apply to current research on extrasolar systems. This sequence deepens understanding of these principles within modern astrophysics, enriching science education. Such activities spark new research questions akin to those debated in scientific circles, enhancing insights into planetary formation.

When German astronomer Johannes Kepler (1571-1630) formulated the three laws of planetary motion, he based his work on Tycho Brahe's observations of our solar System. Now, 400 years later, we know about exoplanets, or planets outside our solar System. The first exoplanet was detected in 1992[2], and since then, over 5000 exoplanets have been confirmed in more than 4000 planetary systems[3].

Although scientific knowledge is constantly expanding, high school science education often focuses on specific content and may not always keep up with the latest developments in astrophysics. In the specific case of Chile, although it is a geographically privileged country for astronomical research, there is a scarcity of astronomy content across all school levels[4,5], a situation that is repeated internationally[5]. Several authors agree that using astronomy as a context to apply physical laws can be much more engaging in promoting learning[6,7]. In fact, some authors have recognized the need to incorporate these new research topics into the curriculum and have proposed activities that involve the use of real data[8-11]. According to these authors, the real data refers to those data. In this way, activities that involve real data can help close the misalignment between classroom learning and current developments in the field[12,13].

Planetary formation has emerged as an active and cutting-edge field, driven by recent scientific advancements. The discovery of thousands of exoplanets in recent decades has ignited significant interest in understanding how planetary systems form and evolve across various stellar environments. Recent research suggests that planetary formation may exhibit a degree of "uniformity," akin to a "peas in a pod" model[14]. Studies have focused on the similarities in radii of the planetary orbits, masses, densities, and period ratios within planetary systems, indicating greater uniformity in radius and density. Additionally, a transition in system uniformity based on planet mass and radius has been observed. Moreover, the presence of compact, orderly planetary systems with planets of consistent size, regular orbital spacing, and low eccentricities has been highlighted[15]. While these studies remain inconclusive, they prompt intriguing classroom discussions about the diversity of planetary systems, their formation[16,17], and the potential for uniformity as a common characteristic, reflecting ongoing scientific inquiries.

In this project, we outline a teaching sequence using real exoplanet data to teach high school students about Kepler's third law and Newton's law of universal gravitation. The approach focuses on data analysis through plotting and exploring relevant questions, avoiding complex equations. Activities include using an Excel spreadsheet with orbital period, semimajor axis, and mass data (in years, astronomical units, and solar masses, respectively) for our solar system and 59 other extrasolar systems. Data is sourced from the Open Exoplanet Catalog[18] and supplemented with NASA Eyes on Exoplanets data where necessary, both freely available. Selected planetary systems meet criteria of hosting at least two planets, orbiting a single star, and providing complete orbital information (mass, period, semimajor axis). Ideally, groups of three students share one computer for activities, totaling about 3 hours.

The next section provides guiding questions for teachers to facilitate classroom activities, detailing the analyses conducted. Finally, implications of the approach are discussed in the Conclusions.

## Teaching Sequence

In 1619, Kepler proposed a harmonic relationship between the square of the planets' orbital periods and the cube of their average distance from the Sun, based on data from only six known planets at the time. The well-known Kepler's third law can be expressed mathematically as

$$T^2 = k\, a^3 \qquad (1)$$

where T is the orbital period and a is the semimajor axis of the planet's orbit.

Students create a graph relating the cube of the semimajor axis (in astronomical units) and the square of the orbital period (in years) using solar system data. They observe that the proportionality constant k is the slope, with a value of 1 $year^2/AU^3$, using Earth's orbit as a reference. A year is the time it takes for Earth to orbit the Sun, and 1 AU is the average distance between Earth and the Sun, approximately 150 million kilometers. This exercise, common in physics classes, is enhanced with carefully selected questions (Q) to deepen understanding and promote interactive learning.

**Q1.** Do you think other planetary systems have a proportional relationship between the period and semimajor axis similar to that of our solar system?

To answer this, each group receives data from different planetary systems and replicates the solar system analysis. Since Chilean students do not yet know logarithmic functions, they graph by cubing or squaring the data table values, as shown in Fig. 1 for different planetary systems.

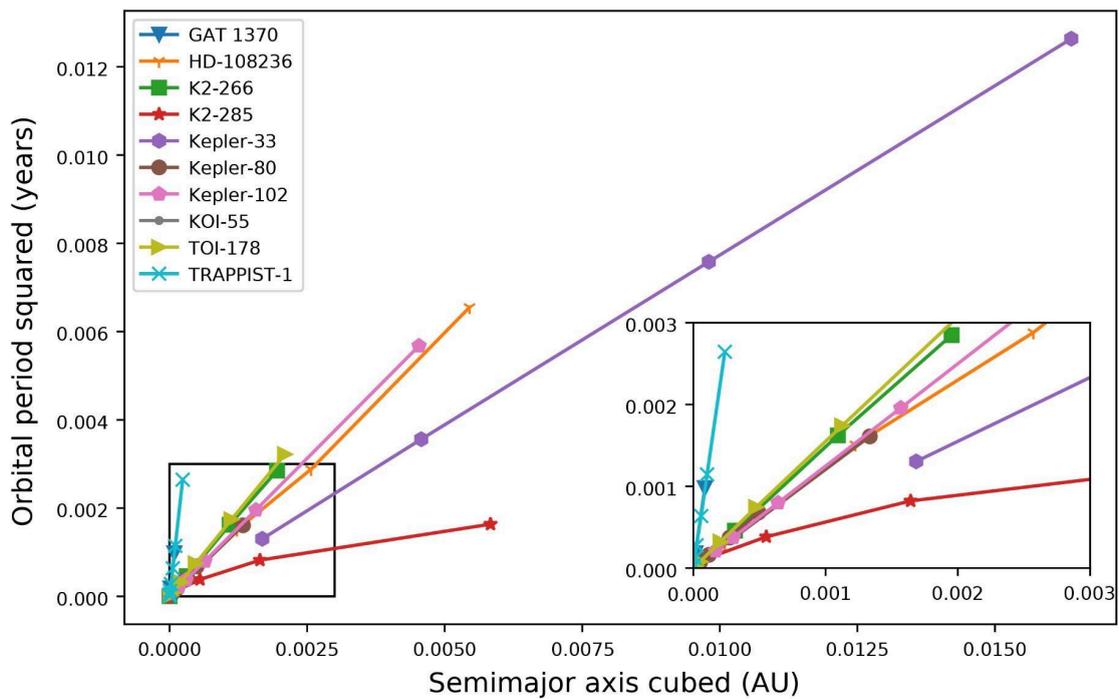

*Fig. 1. Relationship between the squared period and the cubed semi-axis for different extrasolar systems, demonstrating a linear relationship. The inset plot provides a closer look of the linear trend, in order to see more detail of the systems with small orbits.*

Similar to the solar system analysis, students can use the plot to obtain the constant k from the slope. Examples of k values for each extrasolar system are in Table I.

Table I: Value of k obtained from analyzing the slope of Kepler's third law for exoplanetary systems orbiting a single star (see Fig. 1).

| Exoplanetary System | k ($year^2/AU^3$) | Exoplanetary System | k ($year^2/AU^3$) |
| --- | --- | --- | --- |
| GAT 1370 | 11,27 | Kepler-80 | 1,38 |
| HD-108236 | 1,19 | Kepler-102 | 1,25 |
| K2-266 | 1,47 | KOI-55 | 1,85 |
| K2-285 | 0,78 | TOI-178 | 1,56 |
| Kepler-33 | 0,77 | TRAPPIST-1 | 10,93 |

The teacher can emphasize that the unit of the constant of proportionality k was defined for convenience by considering Earth as a reference. Discussing the meaning of this value or its units for other planetary systems can lead to interesting classroom discussions from an epistemological perspective.

To continue the activity and extend the discussion, the teacher can apply Kepler's third law to either the sample of the 10 extrasolar systems in Table I or the total sample composed of 59 exoplanetary systems, presented in Table V in the supplementary material. Regardless of the sample size, it is interesting to note a certain inclination toward values in the range between 1 and 2, regardless of the accepted unit convention discussed above. Nevertheless, some systems, such as TRAPPIST-1 and GAT 1370, exhibit considerably higher k values than those observed in other extrasolar systems.

This remains an open question that can be answered with information provided in the next activity. An alternative approach, assuming students are familiar with logarithms, is to use Kepler's third law expressed as:

$$\log(T) = \frac{3}{2}\log(a) + \frac{1}{2}\log(k) \qquad (2)$$

This analysis shows that for most planetary systems studied, the period and semimajor axis are close to the 3:2 ratio determined by Kepler. Table II compares the slope obtained by Eq. (2) for the same 10 systems represented in Fig. 1 and Table VI for the total sample in the supplementary material. From Table II, it can be seen that K2-285 and KOI-55 differ from the rest, with values slightly away from the 3:2 ratio.

Table II. Value of the slope obtained for exoplanetary systems orbiting a single star [using the method suggested by Eq. (2)].

| Exoplanetary System | Slope | Exoplanetary System | Slope |
|---|---|---|---|
| GAT 1370 | 1,49 | Kepler-80 | 1,49 |
| HD-108236 | 1,49 | Kepler-102 | 1,50 |
| K2-266 | 1,50 | KOI-55 | 1,21 |
| K2-285 | 0,94 | TOI-178 | 1,50 |
| Kepler-33 | 1,49 | TRAPPIST-1 | 1,52 |

On one hand, a review of the literature reveals that KOI- 55 is a peculiar system. It is a compact system of two planets orbiting near an evolved post-giant star. This star has undergone a red giant phase, during which the star's core no longer contains hydrogen. This results in the star's expansion and the fusion of hydrogen from layers near the core. During this phase, the star may engulf planets in its vicinity.

Additionally, the core reaches a sufficient temperature to fuse helium to carbon, a process known as triple alpha. At this stage, the star becomes unstable, resulting in fluctuations in brightness due to the expansion and contraction of its interior[19]. In the case of KOI-55, it is hypothesized that these two planets were able to survive this stage[20]. However, a subsequent study suggested that the signals associated with the detection of both planets were instead due to pulsations of the red subgiant star.[21] This red giant phase probably modified the architecture of the system in a way that makes it hard to interpret in terms of Kepler's laws. This is precisely an opportunity to introduce in the classroom a discussion around limitations in physical laws.

Table III. Examples of solar masses obtained by applying Newton's law of universal gravitation and Kepler's third law to the planet b in some selected extrasolar systems.

| Planet name | Star mass obtained (Solar Mass) |
|---|---|
| GAT 1370 | 0,09 |
| HD-108236 | 0,84 |
| K2-266 | 0,68 |
| K2-285 | 1,28 |
| Kepler-33 | 1,30 |
| Kepler-80 | 0,72 |
| Kepler-102 | 0,80 |
| KOI-55 | 0,54 |
| TOI-178 | 0,67 |
| TRAPPIST-1 | 0,09 |

Table IV. Values of k and slope of Eqs. (2) and (4) for two P-type extrasolar systems.

| Planetary System | $k(year^2/AU^3)$ | slope |
|---|---|---|
| NN SER (AB) | 1,53 | 1,49 |
| Kepler-47 | 0,71 | 1,51 |

On the other hand, the K2-285 system consists of a main sequence star (a star that obtains its energy through hydrogen fusion in its core) and four exoplanets, whose orbital parameters were determined by the transit and radial velocity methods. The orbital periods of the exoplanets are close to a 1:2:3:4 ratio, suggesting that the system may be in resonance.[22] This highlights the incomplete nature of research on astronomical systems, which can lead to revisions of orbital parameters and a significant margin of uncertainty in the associated data.

**Q2.** Do all stars that host planetary systems necessarily have the same mass as the Sun?

Currently, astronomers apply Kepler's third law alongside Newton's law of universal gravitation to determine parameters of newly discovered exoplanet systems. According to Newton's law, gravitational attraction between two objects is proportional to the product of their masses and inversely proportional to the square of the distance between them.

$$\vec{F} = \frac{-GMm}{r^2}\hat{r} \qquad (3)$$

where G is the gravitational constant equivalent to $6{,}67 \cdot 10^{-11}\ m^3/(kg \cdot s^2)$, M is the mass of the star, and m is the mass of the planet.

Assuming the planets move in circular orbits with a certain speed around the host star, and neglecting the mass of the planet compared to the star, the mass M of the star can be determined using the following expression:

$$M = \frac{4\pi^2}{G}\frac{a^3}{T^2} \qquad (4)$$

In this activity, eight extrasolar systems were selected where the stellar mass can be derived from the data of their b planets, the first discovered in their extrasolar system. Although the exercise appears straightforward, as it only involves the replacement of values in Eq. (3), the challenge lies in the conversion of units, because students must present the result in solar masses. By filling the values in Table III, students will observe that there is a wide range of star masses, thereby providing a definitive answer to the initial question.

To extend the activity, the teacher can focus on the relation between star mass and the value of k. Using Eqs. (1) and (4), k can be calculated. With this information, the teacher can revisit the open question about TRAPPIST-1 and GAT 1370. Table III shows that these stars have masses around 0.09 solar masses. Therefore, students can conclude that k is inversely proportional to the star's mass in the system. Higher k values are expected in systems with lower stellar masses, k values around 1 are expected for Sun-like stars, and lower k values are expected for systems with higher stellar masses.

So far, we have focused on systems with a single host star. However, there are extrasolar systems composed of two stars, known as binary systems. In this context, it's intriguing for teachers to consider Kepler's third law for these cases.

**Q3.** How does Kepler's third law apply to different types of extrasolar systems?

In cases involving more than two interacting objects, Kepler's third law applies to all bodies involved. For example, in binary star systems, astronomers determine the masses of stars and study their evolution and dynamics by observing their orbits. In extrasolar systems with multiple stars, categorized as S-type (orbiting one star while another perturbs the system) and P-type (circumbinary, where planets orbit both stars concurrently[23]), Kepler's law aids in understanding complex gravitational interactions affecting star orbits and evolution.

According to the Open Exoplanet Catalog, complete data is available for only two circumbinary systems: NN SER (AB) and Kepler-47, as detailed in Table IV. It's noteworthy that these systems also exhibit a 3:2 ratio, as indicated by Eq, (4), despite being three-body systems.

Thus, although the n-body problem does not have a general solution, in the cases analyzed the results obtained are consistent with a two-body problem because the stars of the systems are very close to each other. NN Ser consists of a white dwarf and a red dwarf main sequence star (the former is the core of a sun-like star that is exposed after the star has used up all its fuel, and the latter is a very cool, small main sequence star) with a semimajor axis between them smaller than a solar radius[24,25]. On the other hand, the components of Kepler-47 are a Sun-like star and a companion one-third its size with a separation of ～0.08 AU[26]. Thus, the stellar components can be assumed to be a single body, so it makes sense that Kepler's third law applies to these systems.

## Conclusions

The proposed teaching sequence enables classroom discussion of current scientific questions as new data emerges. It is remarkable to find these results easily aligning with what is reported in scientific literature regarding potential uniformity in compact planetary systems[14].

Newton's law of universal gravitation and Kepler's third law should be recontextualized in education to avoid presenting them solely as historical facts.Using real data and simple tools such as Excel can facilitate engaging discussions about the application of fundamental physics laws in cutting-edge astrophysics research. Moreover, integrating this activity across physics and mathematics classes enhances data analysis skills, including the use of logarithmic functions and graph construction. Implementing such activities sparks new research questions in the classroom akin to those debated in scientific circles, fostering a deeper understanding of planetary formation.

One constraint to introducing cutting-edge topics in the classroom is that scientific data from open catalogs or journals are often not readily usable in schools and require adaptation by teachers. In our project, collaborating scientists undertook this task, underscoring the importance of collaborations between educational and scientific communities[12]. In order for this proposal to be implemented, we include in the Supplementary Material a complete list of available data for 59 extrasolar planetary systems.

Finally, we consider that the activities suggested in this work can be an important opportunity to introduce in physics classes interesting discussions from a perspective of the nature of science. The limitations of physical laws and their constant reconsideration as a result of new findings allow us to teach physics as a fascinating and useful discipline, far from a reductionist and decontextualized vision. All the analyses performed and results obtained do not pretend to be conclusive. In contrast, we expected to motivate great conversations and the formulation of new questions in the classroom that the scientific community is seeking to answer.

## Acknowledgments

This work was supported by the ANID Millennium Science Initiative Program under Grant NCN2024_001; Joint Committee ESO–Chile through the CIRAS Pedagogical Team project (2021-2023); and DICYT Regular project – Vicerrectoría de Investigación, Innovación y Creación at the Universidad de Santiago de Chile under Grant 042431HS.

Author Declarations Section

The authors have no conflicts to disclose.

# Bibliography

1. National Science Foundation. (2015). Scientific education in the 21st century. National Science Foundation Report, 25(3), 112-125.
2. Wolszczan, A., & Frail, D. A. (1992). A planetary system around the millisecond pulsar PSR1257+ 12. *Nature*, *355*(6356), 145-147.
3. Lissauer, J. J., Rowe, J. F., Jontof-Hutter, D., Fabrycky, D. C., Ford, E. B., Ragozzine, D., ... & Nizam, K. M. (2024). Updated Catalog of Kepler Planet Candidates: Focus on Accuracy and Orbital Periods. *The Planetary Science Journal,* 5(6), 152
4. Barr, A. Hernández, C. Caerols, H. Arias, M. Iturra, D. (2021). Statement of Astronomy Education in Chile, doi:10.5281/zenodo.6015016
5. Rodrigues, L., Montenegro, M., & Meneses, A. (2023). Mapping the astronomy content knowledge of Chilean in-service teachers. *International Journal of Science Education*, *45*(6), 451-469.
6. Salimpour, S., Bartlett, S., Fitzgerald, M. T., McKinnon, D. H., Cutts, K. R., James, C. R., ... & Ortiz-Gil, A. (2021). The gateway science: A review of astronomy in the OECD school curricula, including China and South Africa. *Research in Science Education*, *51*, 975-996.
7. van Dishoeck, E. F., & Elmegreen, D. M. (2018). The IAU Strategic Plan for 2020-2030: OAO. *Proceedings of the International Astronomical Union*, *14*(A30), 546-548.
8. LoPresto, M. C. (2019). Using real data from the Kepler mission to find potentially habitable planets: An introductory astronomy exercise. *The Physics Teacher*, *57*(3), 159-162.
9. Hyde, J. M. (2021). Exploring Hubble Constant Data in an Introductory Course. *The Physics Teacher*, *59*(3), 159-161.
10. Della-Rose, D., Carlson, R., de La Harpe, K., Novotny, S., & Polsgrove, D. (2018). Exoplanet science in the classroom: Learning activities for an introductory physics course. *The Physics Teacher*, *56*(3), 170-173.
11. Gould, A., Komatsu, T., DeVore, E., Harman, P., & Koch, D. (2015). Kepler's third law and NASA's Kepler mission. *The Physics Teacher*, *53*(4), 201-204.
12. C. Hernández, I. Fuentes-Morales, S. Perez, I. Benito, F. Alarcón (2023). Using authentic exoplanet data to promote active learning of physics and mathematics in schools" Conference proceedings. Doi:10.5281/zenodo.7705416
13. Fuentes Morales, I., Hernández, C., Alarcon, F., Benito, I., & Montecinos, R. (2024). Construction of color-magnitude diagrams using real astronomical data for teaching at school. *European Journal of Physics*. In press. DOI: 10.1088/1361-6404/ad3d44
14. Otegi, J. F., Helled, R., & Bouchy, F. (2022). The similarity of multi-planet systems. *Astronomy & Astrophysics*, *658*, A107.

15. Weiss, L. M., Millholland, S. C., Petigura, E. A., Adams, F. C., Batygin, K., Bloch, A. M., & Mordasini, C. (2023). Architectures of compact multi-planet systems: diversity and uniformity, in *Protostars and Planets VII,* vol. 534, p. 863.
16. Perez, S., Dunhill, A., Casassus, S., Roman, P., Szulágyi, J., Flores, C., ... & Montesinos, M. (2015). Planet formation signposts: observability of circumplanetary disks via gas kinematics. *The Astrophysical Journal Letters*, *811*(1), L5.
17. Weber, P., Pérez, S., Zurlo, A., Miley, J., Hales, A., Cieza, L., ... & Williams, J. (2023). Spirals and clumps in V960 Mon: signs of planet formation via gravitational instability around an FU Ori star?. *The Astrophysical Journal Letters*, *952*(1), L17.
18. https://www.openexoplanetcatalogue.com
19. Carroll, B. W., & Ostlie, D. A. (2017). *An introduction to modern astrophysics*. Cambridge University Press.
20. Charpinet, S., Fontaine, G., Brassard, P., Green, E. M., Van Grootel, V., Randall, S. K., ... & Telting, J. H. (2011). A compact system of small planets around a former red-giant star. *Nature*, *480*(7378), 496-499.
21. Krzesinski, J. (2015). Planetary candidates around the pulsating sdB star KIC 5807616 are considered doubtful. *Astronomy & Astrophysics*, *581*, A7.
22. Palle, E., Nowak, G., Luque, R., Hidalgo, D., Barragán, O., Prieto-Arranz, J., ... & Zechmeister, M. (2019). Detection and Doppler monitoring of K2-285 (EPIC 246471491), a system of four transiting planets smaller than Neptune. *Astronomy & Astrophysics*, *623*, A41.
23. Dvorak, R. (1982). Planetenbahnen in Doppelsternsystemen. *Österreichische Akademie Wissenschaften Mathematisch Naturwissenschaftliche Klasse Sitzungsberichte Abteilung*, *191*(10), 423-437.
24. Parsons, S. G., Marsh, T. R., Copperwheat, C. M., Dhillon, V. S., Littlefair, S. P., Gänsicke, B. T., & Hickman, R. (2010). Precise mass and radius values for the white dwarf and low mass M dwarf in the pre-cataclysmic binary NN Serpentis. *Monthly Notices of the Royal Astronomical Society*, *402*(4), 2591-2608.
25. Beuermann, K., Dreizler, S., & Hessman, F. V. (2013). The quest for companions to post-common envelope binaries-IV. The 2: 1 mean-motion resonance of the planets orbiting NN Serpentis. *Astronomy & Astrophysics*, *555*, A133.
26. Orosz, J. A., Welsh, W. F., Carter, J. A., Fabrycky, D. C., Cochran, W. D., Endl, M., ... & Borucki, W. J. (2012). Kepler-47: a transiting circumbinary multiplanet system. *Science*, *337*(6101), 1511-1514.